\documentstyle[12pt,floats,pra,aps]{revtex}
\tightenlines 
\begin{document} \draft

\title{\LARGE \bf QED in the Presence of Arbitrary Kramers--Kronig
Dielectric Media}
\author{Stefan Scheel and Dirk--Gunnar Welsch}
\address{Theoretisch-Physikalisches Institut, 
Friedrich-Schiller-Universit\"at Jena,\\
Max-Wien-Platz 1,
D-07743 Jena, Germany}

\maketitle
\thispagestyle{empty}
\begin{abstract}
The phenomenological Maxwell field is quantized for
arbitrarily space- and frequency-dependent complex
permittivity. The formalism takes account of the 
Kramers--Kronig relation and the
dissipation-fluctuation theorem and yields the fundamental 
equal-time commutation relations of QED. Applications
to the quantum-state transformation at absorbing and amplifying 
four-port devices and to the spontaneous decay of an excited
atom in the presence of absorbing dielectric bodies are discussed.
\end{abstract}

\vspace{8mm}

\section{Introduction}
\label{intro}

Quantization of the electromagnetic field in dispersive and absorbing
dielectrics requires a concept which is consistent with both the
principle of causality and the dissipation--fluctuation theorem and 
which necessarily yields the fundamental equal-time commutation 
relations of QED. In order to achieve this goal, several approaches
are possible. The microscopic approach starts from the 
exact Hamiltonian of the coupled radiation--matter system and  
integrates out, in some approximation, the matter degrees of 
freedom to obtain an effective theory for the electromagnetic field.
Since the procedure can hardly be performed for arbitrary 
media, simplified model systems are considered. A typical example
is the use of harmonic-oscillator models for the matter polarization 
and the reservoir variables together with the assumption of
bilinear couplings \cite{Huttner}. In the macroscopic approach, 
the phenomenological Maxwell theory, in which the effect of 
the medium is described in terms of constitutive equations, is
quantized. Since this concept does not use any microscopic 
description of the medium, it has the benefit of being 
universally valid, at least as long as the medium can be regarded 
as a continuum. 

Here we study the problem of quantization of the
phenomenological Maxwell theory for nonmagnetic but otherwise
arbitrary linear media at rest, starting from the classical
Green function integral representation of the electromagnetic
field. The method was first established for one-dimensional
systems \cite{Gruner} and simple three-dimensional systems
\cite{Ho} and later generalized to arbitrary inhomogeneous 
dielectrics described in terms of a spatially varying permittivity 
which is a complex function of frequency \cite{Scheel98}. 
 
In Sec.~\ref{quant} we briefly review the quantization scheme and
give an extension to anisotropic dielectrics (including amplifying
media), which complete the class of nonmagnetic (local) media.
In Sec.~\ref{state} we apply the method to the problem of
quantum-state transformation at absorbing and amplifying four-port
devices, and in Sec.~\ref{decay} we give an application to the 
problem of spontaneous decay of an excited atom in the presence of 
absorbing bodies.

\section{Quantization scheme}
\label{quant}

\vspace{-1ex}

Let us first consider the electromagnetic field in
isotropic dielectrics without external sources. The (operator-valued)
phenomenological Maxwell equations in the temporal Fourier space read
\vspace{-1ex}
\begin{eqnarray}
\label{2.1}
{\bf \mbox{\boldmath $\nabla$} } \cdot \hat{\underline{\bf B}}({\bf
r},\omega) = 0, 
& \qquad & {\bf \mbox{\boldmath $\nabla$} } \times \hat{\underline{\bf
E}}({\bf r},\omega) 
= i\omega \hat{\underline{\bf B}}({\bf r},\omega), \\
{\bf \mbox{\boldmath $\nabla$} } \cdot [\epsilon_0 \epsilon({\bf r},\omega)
\hat{\underline{\bf E}}({\bf r},\omega) ] =
\hat{\underline{\rho}}({\bf r},\omega),
& \qquad & {\bf \mbox{\boldmath $\nabla$} } \times \hat{\underline{\bf
B}}({\bf r},\omega) 
=-i (\omega/c^2) \epsilon({\bf r},\omega) \hat{\underline{\bf
E}}({\bf r},\omega) +\mu_0 \hat{\underline{\bf j}}({\bf r},\omega) .
\end{eqnarray}

\vspace{-.5ex}
\noindent
   From the principle of causality it follows that the complex-valued
permittivity $\epsilon({\bf r},\omega)$ $\!=$
$\!\epsilon_R({\bf r},\omega)$ $\!+$ $\!i\,\epsilon_I({\bf r},\omega)$
satisfies the Kramers--Kronig relations. 
Hence, it is a holomorphic function in the upper complex
frequency plane without poles and zeros and approaches unity in the
high-frequency limit. Consistency with the dissipation--fluctuation theorem
requires the introduction of an operator noise charge density
$\hat{\underline{\rho}}({\bf r},\omega)$ and an operator noise current
density $\hat{\underline{\bf j}}({\bf r},\omega)$ satisfying the
equation of continuity. Quantization is performed by introducing
bosonic vector fields $\hat{\bf f}({\bf r},\omega)$,
\vspace{-1ex}
\begin{equation}
\label{2.2}
\hat{\underline{\bf j}}({\bf r},\omega) = 
\omega \sqrt{\hbar\epsilon_0\epsilon_I({\bf r},\omega)/\pi} 
\, \hat{\bf f}({\bf r},\omega) ,
\end{equation}

\vspace{-.5ex}
\noindent
which play the role of the fundamental variables of the theory. All
relevant operators of the system such as the electric and magnetic 
fields and the matter polarization can be constructed in terms of them. 
For example, the operator of the electric field is given by 
the integral representation
\vspace{-1ex}
\begin{equation}
\label{2.3}
\hat{E}_k({\bf r}) 
= i\mu_0 \sqrt{\hbar\epsilon_0/\pi} 
\int_0^\infty d\omega
\int d^3{\bf r'} \,\omega^2 \sqrt{\epsilon_I({\bf r'},\omega)} 
\,G_{kk'}({\bf r},{\bf r'},\omega)
\hat{f}_{k'}({\bf r'},\omega)
+ {\rm H.c.},
\end{equation}

\vspace{-.5ex}
\noindent
with $G_{kk'}({\bf r},{\bf r'},\omega)$ being the classical 
dyadic Green function. This representation
together with the fundamental relation
\vspace{-.5ex}
\begin{equation}
\label{4.3}
\int d^3{\bf s} \, (\omega/c)^2 \epsilon_I({\bf s},\omega)
G_{ik}({\bf r},{\bf s},\omega) G^\ast_{jk}({\bf r'},{\bf s},\omega) =
{\rm Im}\, G_{ij}({\bf r},{\bf r'},\omega),
\end{equation}

\vspace{-.5ex}
\noindent
which follows directly from the partial differential equation for the
dyadic Green function, leads to the equal-time 
commutation relation \cite{Ho}
\vspace{-.5ex}
\begin{equation}
\label{2.4}
\left[ \epsilon_0 \hat{E}_k({\bf r}), \hat{B}_l({\bf r'}) \right] =
(\hbar/\pi) \, \epsilon_{lmk'} \partial^{r'}_m
\int_{-\infty}^\infty d\omega \, (\omega/c^2) 
G_{kk'}({\bf r},{\bf r'},\omega) .
\end{equation}

\vspace{-.5ex}
\noindent
Using general properties of the Green function, it can be shown 
\cite{Scheel98} that Eq.~(\ref{2.3}) reduces, 
for arbitrary $\epsilon({\bf r},\omega)$, to the well-known QED 
commutation relation
\vspace{-.5ex}
\begin{equation}
\label{2.5}
\left[ \epsilon_0 \hat{E}_k({\bf r}), \hat{B}_l({\bf r'}) \right] = 
i\hbar \epsilon_{klm} \partial_m^{r'} \delta({\bf r}-{\bf r'}) 
\end{equation}

\vspace{-.5ex}

The extension to anisotropic and amplifying media is
straightforward, since we may assume the medium to be reciprocal, 
so that the permittivity tensor $\epsilon_{ij}({\bf r},\omega)$ is
necessarily symmetric. In particular, $\epsilon_{ij}({\bf r},\omega)$ 
can be diagonalized by an orthogonal matrix $O_{kl}({\bf r},\omega)$. 
With regard to amplifying media, we note that amplification 
requires the role of the noise creation and annihilation operators 
to be exchanged. The calculation then shows that the fundamental 
relation (\ref{2.2}) can be generalized to
\vspace{-.5ex}
\begin{equation}
\label{2.6}
\hat{\!\underline{j}}_i({\bf r},\omega) = \omega
\sqrt{\hbar\epsilon_0/\pi} \left[ \gamma_{ij}^-({\bf
r},\omega) \hat{f}_j({\bf r},\omega) +\gamma_{ij}^+({\bf r},\omega)
\hat{f}^\dagger_j({\bf r},\omega) \right] ,
\end{equation}
with
\vspace{-2.5ex}
\begin{equation}
\label{2.7}
\gamma_{ij}^\mp({\bf r},\omega) = O_{ik}({\bf r},\omega) \sqrt{\left|
\tilde{\epsilon}_{kl\;I}({\bf r},\omega) \right|} \; O_{lj}^{-1}({\bf
r},\omega) \;\Theta\left[\pm \tilde{\epsilon}_{kl\;I}({\bf r},\omega)
\right],
\end{equation}
\vspace{-4.5ex}
\begin{equation}
\label{2.8}
\tilde{\epsilon}_{ij\;I}({\bf r},\omega) = \delta_{ij}
\epsilon^{(i)}_I({\bf r},\omega) = O_{ik}^{-1}({\bf r},\omega)
\epsilon_{kl\;I}({\bf r},\omega) O_{lj}({\bf r},\omega) .
\end{equation}
Equation (\ref{2.6}) completes the quantization scheme for the
electromagnetic field in arbitrary linear, nonmagnetic (local) media.

\section{Quantum-state transformations by absorbing and amplifying
four-port devices}
\label{state}

\vspace{-1ex}

Let us first apply the theory to the problem of quantum-state 
transformation at absorbing and amplifying four-port devices 
such as beam-splitter-like devices.
Specifying the formulas to the one-dimensional case for simplicity
and rewriting the integral representation (\ref{2.5}) in terms of 
amplitude operators $\hat{a}_j(\omega)$ and $\hat{b}_j(\omega)$ for 
the incoming and outgoing waves ($j$ $\!=$ $\!1,2$), the action 
of an absorbing device can be given by the (vector)
operator transformation
\vspace{-1ex}
\begin{equation}
\label{3.1}
\hat{\bf b}(\omega) = {\bf T}(\omega) \hat{\bf a}(\omega) +{\bf
A}(\omega) \hat{\bf g}(\omega),
\end{equation}

\vspace{-1ex}
\noindent
where $\hat{g}_j(\omega)$ are the operators of device excitations and
${\bf T}(\omega)$ and ${\bf A}(\omega)$ are the characteristic
transformation and absorption matrices of the device given in terms of
its complex refractive-index profile \cite{Gruner2}. Note that
$\hat{a}_j(\omega)$ and $\hat{g}_j(\omega)$ are independent bosonic
operators. Further, it can be shown that the relation
${\bf T}(\omega) {\bf T}^+(\omega)$ $\!+$ $\!{\bf A}(\omega) 
{\bf A}^+(\omega)$ $\!=$ $\!{\bf I}$ is satisfied, which
ensures bosonic commutation relations for $\hat{b}_j(\omega)$. In
order to construct the unitary transformation, we introduce some 
auxiliary (bosonic) device variables $\hat{h}_j(\omega)$, 
combine the two-vectors
$\hat{\bf a}(\omega)$ and $\hat{\bf g}(\omega)$ to the four-vector
$\hat{\mbox{\boldmath $\alpha$}}(\omega)$, and accordingly
$\hat{\bf b}(\omega)$ and $\hat{\bf h}(\omega)$ to 
$\hat{\mbox{\boldmath $\beta$}}(\omega)$. The four-vectors 
$\hat{\mbox{\boldmath $\alpha$}}(\omega)$ and 
$\hat{\mbox{\boldmath $\beta$}}(\omega)$ are related to each other as
\vspace{-1ex}
\begin{equation}
\label{3.2}
\hat{\mbox{\boldmath $\beta$}}(\omega) = \mbox{\boldmath
$\Lambda$}(\omega) \hat{\mbox{\boldmath $\alpha$}}(\omega), \qquad
\mbox{\boldmath $\Lambda$}(\omega) \in \mbox{SU(4)}.
\end{equation}

\vspace{-1ex}
\noindent
Introducing the positive Hermitian matrices ${\bf C}(\omega)$
$\!=$ $\!\sqrt{{\bf T}(\omega) {\bf T}^+(\omega)}$ and ${\bf
S}(\omega)$ $\!=$ $\!\sqrt{{\bf A}(\omega) {\bf A}^+(\omega)}$, the
four-matrix $\mbox{\boldmath $\Lambda$}(\omega)$ can be
written in the form \cite{Knoll99}
\vspace{-1ex}
\begin{equation}
\label{3.3}
\mbox{\boldmath $\Lambda$}(\omega) = \left(
\begin{array}{cc}
{\bf T}(\omega) & {\bf A}(\omega) \\
-\lambda{\bf S}(\omega) {\bf C}^{-1}(\omega) {\bf T}(\omega) &
{\bf C}(\omega) {\bf S}^{-1}(\omega) {\bf A}(\omega) 
\end{array} \right)
\end{equation}

\vspace{-1ex}
\noindent
($\lambda$ $\!=$ $\!1$).
The input--output relation (\ref{3.2}) can then be expressed in terms
of a unitary operator transformation $\hat{\mbox{\boldmath
$\beta$}}(\omega)$ $\!=$ $\!\hat{U}^\dagger \hat{\mbox{\boldmath
$\alpha$}}(\omega) \hat{U}$. Equivalently, $\hat{U}$ can be applied to
the density operator of the input quantum state $\hat{\rho}_{\rm in}
[\hat{\mbox{\boldmath $\alpha$}}(\omega), \hat{\mbox{\boldmath
$\alpha$}}^\dagger(\omega)]$, and tracing over the device
variables yields
\vspace{-.5ex}
\begin{equation}
\label{3.4}
\hat{\rho}_{\rm out}^{({\rm Field})} = {\rm Tr}^{({\rm Device})}
\left\{ \hat{\rho}_{\rm in}\!\left[ \mbox{\boldmath $\Lambda$}^+(\omega)
\hat{\mbox{\boldmath $\alpha$}}(\omega), \mbox{\boldmath
$\Lambda$}^T(\omega) \hat{\mbox{\boldmath $\alpha$}}^\dagger(\omega)
\right] \right\} .
\end{equation}

\vspace{-.5ex}
\noindent
To give an example, let us consider the case when one input channel is
prepared in an $n$-photon Fock state and the device and the second
input channel are left in vacuum, i.e., $\hat{\rho}_{\rm in}$ $\!=$
$\!|n,0,0,0\rangle\langle n,0,0,0|$. Applying Eq.~(\ref{3.4}),
after some algebra we derive for the density operator of the $i$-th
output channel
\vspace{-.5ex}
\begin{equation}
\label{3.5}
\hat{\rho}_{{\rm out},i}^{({\rm Field})} = \sum\limits_{k=0}^n {n
\choose k} |T_{i1}|^{2k} \left( 1-|T_{i1}|^2\right)^{n-k} |k\rangle
\langle k| .
\end{equation}

\vspace{-.5ex}
\noindent
Next, let us assume that the two input channels are prepared in single-photon
Fock states, i.e., $\hat{\rho}_{\rm in}$ $\!=$ $\!|1,1,0,0\rangle\langle
1,1,0,0|$. We derive for the density operator of
the $i$-th output channel
\vspace{-.5ex}
\begin{eqnarray}
\label{3.6}
\hat{\rho}_{{\rm out},i}^{({\rm Field})} &=& \left[ 1-|T_{i1}|^2
(1-|T_{i2}|^2) -|T_{i2}|^2 (1-|T_{i1}|^2) \right] |0 \rangle \langle
0| \nonumber \\ &&
+\left( |T_{i1}|^2+ |T_{i2}|^2-4 |T_{i1}|^2 |T_{i2}|^2 \right) |1
\rangle \langle 1| +2 |T_{i1}|^2 |T_{i2}|^2 |2 \rangle \langle 2| .
\end{eqnarray}

\vspace{-.5ex}
\noindent

The extension to amplifying devices is straightforward. One has to
replace the annihilation operators $\hat{g}_j(\omega)$ in
Eq.~(\ref{3.1}) by the corresponding creation operators
$\hat{g}^\dagger_j(\omega)$. This leads again to an
input-output relation of the form (\ref{3.2}) but with 
$\lambda$ $\!=$ $\!-1$ in Eq.~(\ref{3.3}), the matrix
$\mbox{\boldmath $\Lambda$}(\omega)$ being now an element of the
noncompact group SU(2,2).

\section{Spontaneous decay near dielectric bodies}
\label{decay}

\vspace{-1ex}

Spontaneous decay of an excited atom is a process that is
directly related to the quantum vacuum noise, which in the
presence of absorbing bodies is drastically
changed and so is the rate of spontaneous decay, because
of the additional noise introduced by absorption. To
study a radiating (two-level) atom in the presence of
dielectric media, we start from the following Hamiltonian 
in dipole and rotating wave approximations:
\vspace{-1ex}
\begin{equation}
\label{4.1}
\hat{H} = \int d^3{\bf r} \int_0^\infty d\omega \, \hbar \omega
\,\hat{\bf f}^\dagger({\bf r},\omega) \cdot \hat{\bf f}({\bf r},\omega)
+\sum\limits_{\alpha =1}^2 \hbar \omega_\alpha \hat{A}_{\alpha \alpha}
-\left[ i\omega_{21} \hat{A}_{21} \hat{\bf A}^{(+)}({\bf r}_A) \cdot
{\bf d}_{21} +{\rm H.c.} \right] .
\end{equation}

\vspace{-1ex}
\noindent
Here, the atomic operators $\hat{A}_{\alpha \alpha'}$ $\!=$ $\!|\alpha
\rangle\langle \alpha'|$ are introduced, and $\hat{\bf A}^{(+)}({\bf r}_A)$
is the (positive-frequency part of the) vector potential (in Weyl gauge) 
at the position of the atom. Note that the first term in Eq.~(\ref{4.1}) 
is the (diagonal) Hamiltonian of the system that consists of the 
electromagnetic field and the medium (including the dissipative system) 
and is expressed in terms of the fundamental variables 
$\hat{\bf f}({\bf r},\omega)$. Solving the resulting equations
of motion in Markov approximation, the well-known Bloch
equations for the atom are recognized, where the decay rate is 
given by \cite{Scheel99a}
\vspace{-.5ex}
\begin{equation}
\label{4.2}
\Gamma = 2\omega_A^2 \mu_k \mu_{k'}/(\hbar \epsilon_0 c^2) 
\,{\rm Im}\,G_{kk'}({\bf r}_A,{\bf r}_A,\omega_A)
\end{equation}

\vspace{-.5ex}
\noindent
[$\mu_k$ $\!\equiv$ $\!(d_{21})_k$, $\omega_A$ $\!\equiv$
$\!\omega_{21}$]. Note that from Eq.~(\ref{2.3}) together 
with Eq.~(\ref{4.3}) it follows that
\vspace{-.5ex}
\begin{equation}
\label{4.4}
{\rm Im}\, G_{kk'}({\bf r},{\bf r'},\omega) \delta(\omega-\omega') =
\pi \epsilon_0 c^2/(\hbar \omega^2) \langle 0 | \big[
\hat{\underline{E}}_k({\bf r},\omega),
\hat{\underline{E}}^\dagger_{k'}({\bf r'},\omega') \big] | 0 \rangle
\end{equation}

\vspace{-.5ex}
\noindent
in full agreement with the dissipation-fluctuation theorem.

Equation (\ref{4.2}) is valid for any absorbing dielectric 
body. For example, when the atom is sufficiently near to an 
absorbing planar interface, then purely nonradiative decay 
is observed, with \cite{Scheel99b}   
\vspace{-.5ex}
\begin{equation}
\Gamma = \Gamma_0 \left(1 + \frac{\mu_z^2}{\mu^2}\right)
\frac{\epsilon_I(\omega_A)}{
|\epsilon(\omega_A) + 1|^2} \frac{3c^3}{(2\omega_A z)^3}\,, 
\end{equation}

\vspace{-.5ex}
\noindent
where $z$ is the distance between the atom and the interface, and
$\Gamma_0$ is the spontaneous emission rate in free space (for
a guest atom embedded in an absorbing dielectric, see \cite{Scheel99a}).

\vspace{2ex}
\noindent
{\bf Acknowledgement}\\
This work was supported by the Deutsche Forschungsgemeinschaft.

\end{document}